\documentclass{IEEEtran}
\usepackage{cite}
\usepackage{amsmath,amssymb,amsfonts}
\usepackage{graphicx}
\usepackage{textcomp,nicefrac}
\def\BibTeX{{\rm B\kern-.05em{\sc i\kern-.025em b}\kern-.08em
T\kern-.1667em\lower.7ex\hbox{E}\kern-.125emX}}
\begin{document}

\title{High-speed single-photoelectron detection for Cherenkov astronomy

\author{Luca Giangrande, Matthieu Heller, Teresa Montaruli} 
\thanks{This work was supported by the Swiss National Foundation FLARE grant 20FL21 201539.}
\thanks{L. Giangrande, is with Département de Physique Nucléaire et Corpusculaire, Faculté de Science, University of Geneva
(e-mail: luca.giangrande@unige.ch).}
\thanks{M. Heller, is with Département de Physique Nucléaire et Corpusculaire, Faculté de Science, University of Geneva
(e-mail: matthieu.heller@unige.ch).}
\thanks{T. Montaruli, is with Département de Physique Nucléaire et Corpusculaire, Faculté de Science, University of Geneva
(e-mail: teresa.montaruli@unige.ch).}}
\maketitle

\begin{abstract}
Silicon photomultipliers are increasingly replacing photomultiplier tubes in Cherenkov telescope cameras, but achieving single-photoelectron resolution with nanosecond timing in a low-noise, scalable detector system remains challenging. We present a co-designed SiPM sensor and front-end application specific integrated circuit (ASIC) that meets these requirements. The custom hexagonal sensor, developed with Hamamatsu Photonics, incorporates an integrated optical filter and fourfold pixel segmentation. The readout is performed by a second prototype of the FANSIC ASIC, optimized for this application and fabricated in 65~nm standard CMOS technology, it provides eight channels with on-chip analog summing of sub-channels on a $3.5\times 3.5~\mathrm{mm}^2$ die, while consuming only 24~mW per channel.

We demonstrate clear single-photoelectron peak separation with a gain of $2.7 \times 10^{-12}~ \mathrm{V \cdot s}$ , and an impulse response below 4~ns full width at half maximum with a 1.7 ns rise time, preserving the nanosecond-scale structure of Cherenkov pulses. The system responds linearly from 1 to 130 photoelectrons, and 55 distinct photoelectron peaks are resolved by varying the source intensity.

These results demonstrate that the integrated sensor-electronics architecture delivers the speed, resolution, and dynamic range required for imaging atmospheric Cherenkov telescopes, and provides a scalable path toward large-area camera modules.
\end{abstract}


\section{Summary}
\subsection{Scope}
This work aimed to develop and characterize an integrated photodetector system consisting of a custom silicon photomultiplier sensor (SiPM) and a dedicated front-end application-specific integrated circuit (ASIC), co-optimized to achieve single-photoelectron resolution, ultra-fast timing, and stable operation over the dynamic range relevant for ground-based gamma-ray astronomy with imaging atmospheric Cherenkov telescopes (IACT).


\subsection{Background}
Silicon photomultipliers have become the photodetector of choice for next-generation IACT cameras, offering higher photon detection efficiency than conventional photomultiplier tubes, insensitivity to magnetic fields, and lower operating voltages \cite{CTA}. However, their adoption introduces challenges: higher dark count rates, increased sensitivity to the night-sky background, and the need for fast analog signal processing to preserve the nanosecond-scale structure of Cherenkov pulses \cite{FACT}. Several front-end ASICs have been developed for SiPM readout in astroparticle physics, including CITIROC \cite{CITIROC}, PETsys TOFPET \cite{TOFPET}, and the TARGET series \cite{TARGET}. These front ends typically interface with commercial off-the-shelf sensors. The presented work pursues an alternative approach: co-designing the sensor and front-end electronics as a single optimized detector system, including an optical filter integrated at the sensor level by the manufacturer. This strategy enables performance optimizations that are difficult to achieve with independently sourced components.
\subsection{Methods}

\subsubsection{Light sensor}
A hexagonal SiPM sensor based on $50~\mathrm{\mu m}$ cells was developed in collaboration with Hamamatsu Photonics, specifically designed for the focal-plane geometry of imaging atmospheric Cherenkov telescopes. Each pixel is divided into four electrically independent quarters, each with an active area of $23.375~\mathrm{mm}^2$ and a capacitance of approximately 1~nF. Each quarter is connected to a dedicated input channel of the ASIC. An optical filter that suppresses night-sky background light above 450~nm was deposited directly on the sensor, removing the need for external filter coatings at the camera level.

\subsubsection{Front-end electronics}
The presented ASIC is the second prototype of a camera front-end \cite{FANSIC}. The chip provides eight independent readout channels, operates from a single 1.2 V supply, and consumes less than 24~mW of power per pixel. It was fabricated in a standard 65~nm CMOS technology and measures $3.5\times 3.5~\mathrm{mm}^2$. To preserve the fast signal components, the chip is directly wire-bonded onto the front-end printed circuit board, minimizing parasitic capacitances and inductances. The four sub-channels corresponding to one complete pixel are actively summed on-chip, reducing communication overhead while retaining the signal-to-noise benefits of sensor segmentation at the analog level.

\subsubsection{Characterization}
Laboratory measurements were performed using a pulsed monochromatic light source at 375~nm. The sensor was operated at a nominal overvoltage of 6~V. Waveforms were digitized and analyzed to extract gain, timing parameters, single-photoelectron resolution, and linearity across the 1–130 photoelectron occupancy range.

\subsection{Results}
\begin{figure}
    \centering
    \includegraphics[width=0.3\linewidth]{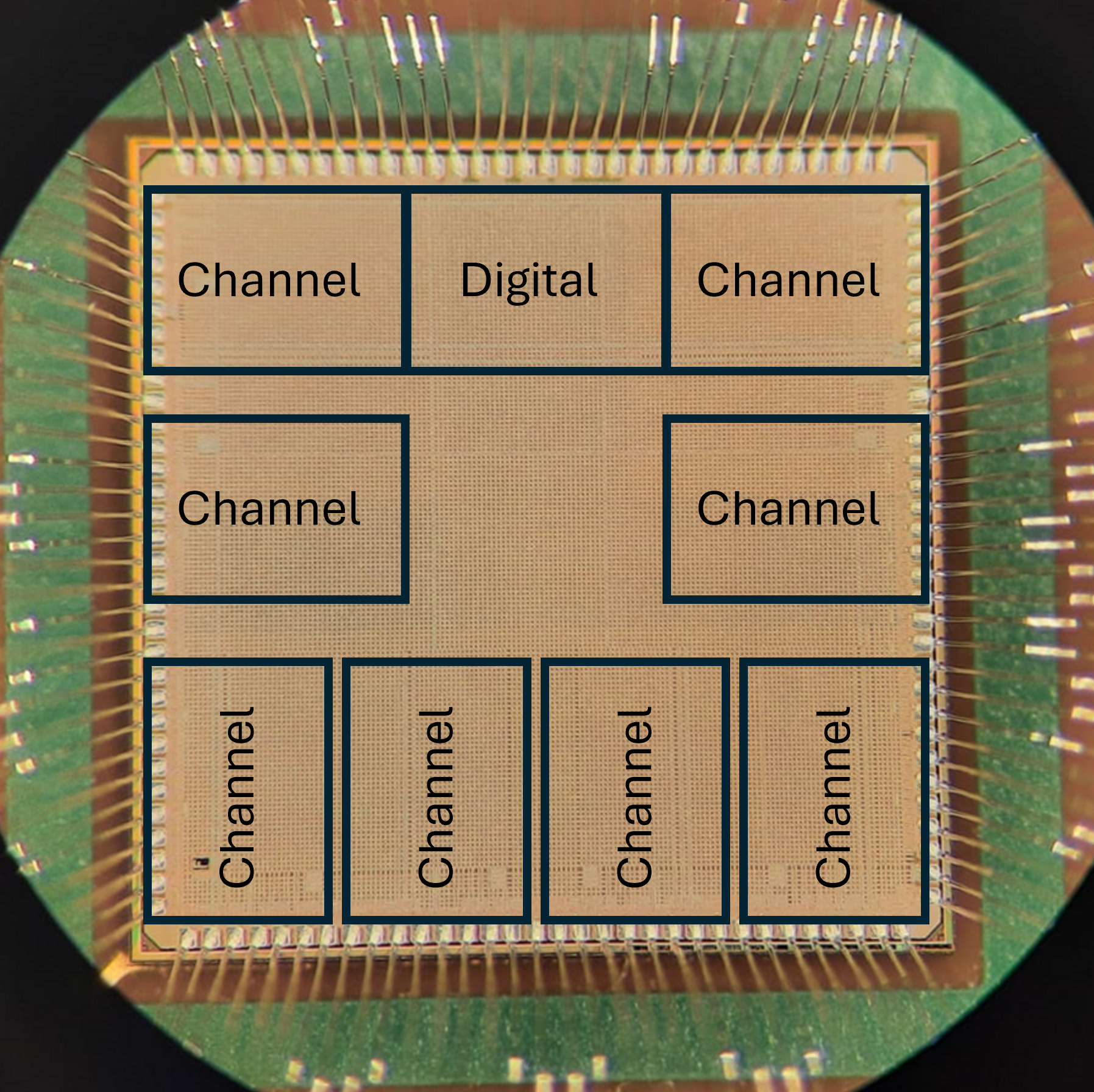}
    \caption{Optical micrograph of the ASIC comprising eight analog channels and one common digital configuration block.}
    \label{fig:micrograph}
\end{figure}

\begin{figure}
    \centering
    \includegraphics[width=0.75\linewidth]{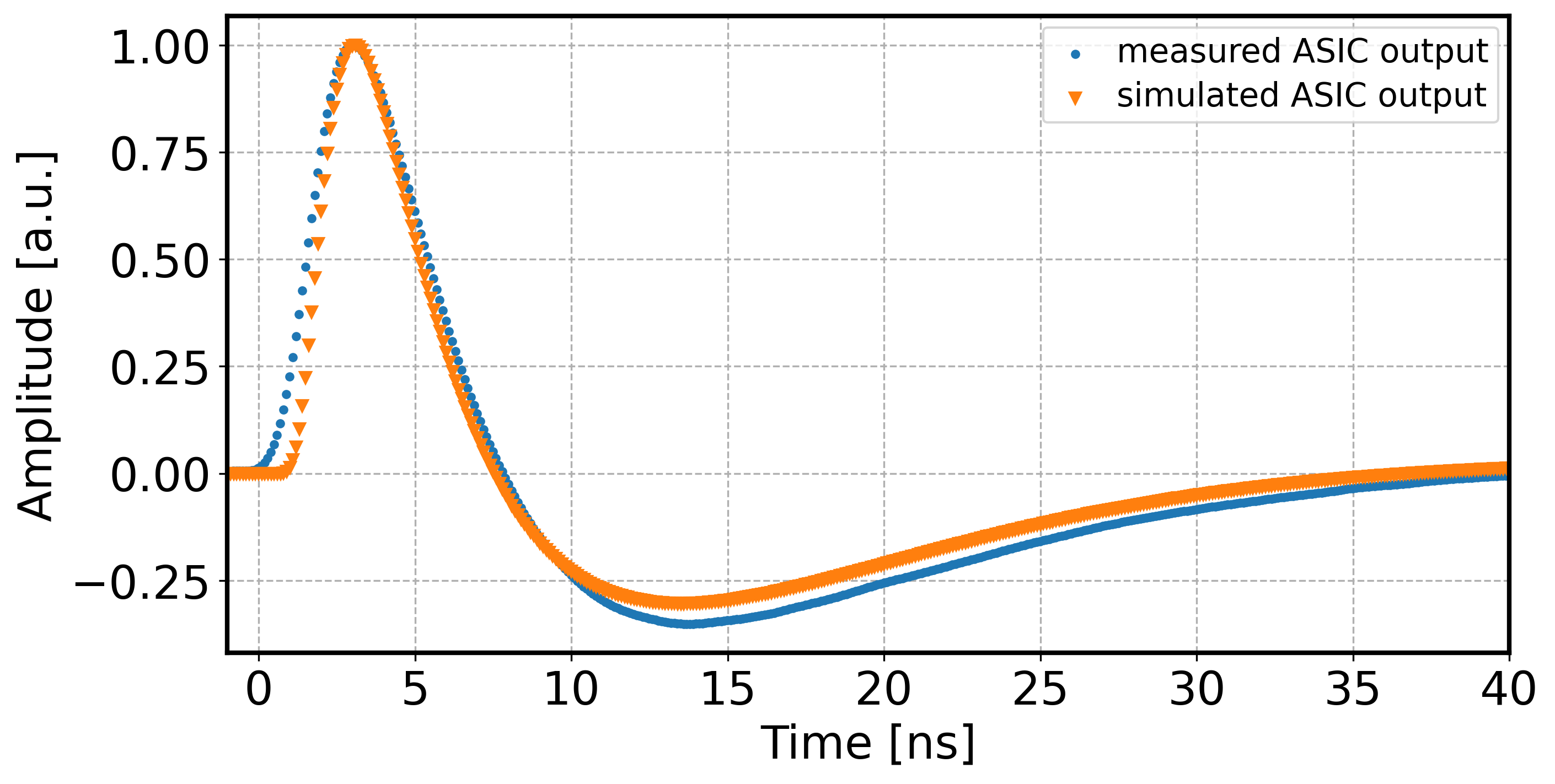}
    \caption{Comparison of the measured and simulated output pulse. The measured waveform represents an average over 10,000 acquisitions. The simulation employs the Corsi model parameterized for the custom sensor.}
    \label{fig:Output_pulse}
\end{figure}

Fig.~\ref{fig:micrograph} shows an optical micrograph of the FANSIC die wire-bonded to its evaluation PCB. The measured impulse response in Fig.\ref{fig:Output_pulse} exhibits a 10\%--90\% rise time of 1.7~ns and a full width at half maximum (FWHM) of 3.99~ns. The recovery time from falling 50\% of peak amplitude to baseline is 32.7~ns.

\begin{figure}
    \centering
    \includegraphics[width=0.75\linewidth]{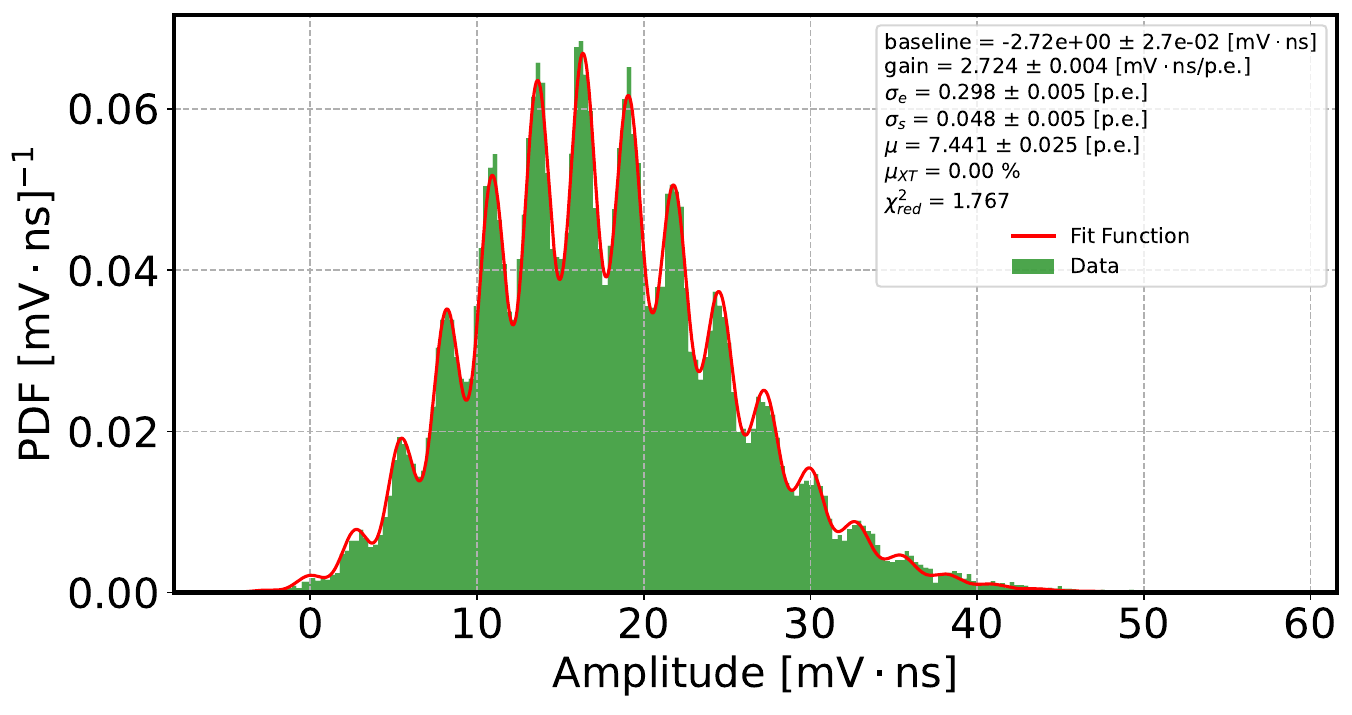}
    \caption{Measured multi‑photoelectron distribution acquired at a fixed light intensity. The distribution is derived from the integral of the output pulse over a fixed window of 4~ns. The data fit employs a generalized Poisson $pdf$ and its parameters are reported in the legend.}
    \label{fig:finger_plot}
\end{figure}

Single-photoelectron resolution is demonstrated in Fig.~\ref{fig:finger_plot}, which shows a multi-photoelectron amplitude spectrum acquired at a mean occupancy of approximately 6 photoelectrons. The histogram contains 60,000 samples. Individual photoelectron peaks are clearly resolved, and the distribution is well described by a generalized Poisson function. The gain derived from the fit corresponds to about $2.7\cdot10^{12}~\mathrm{V~s}$.

\begin{figure}
    \centering
    \includegraphics[width=0.8\linewidth]{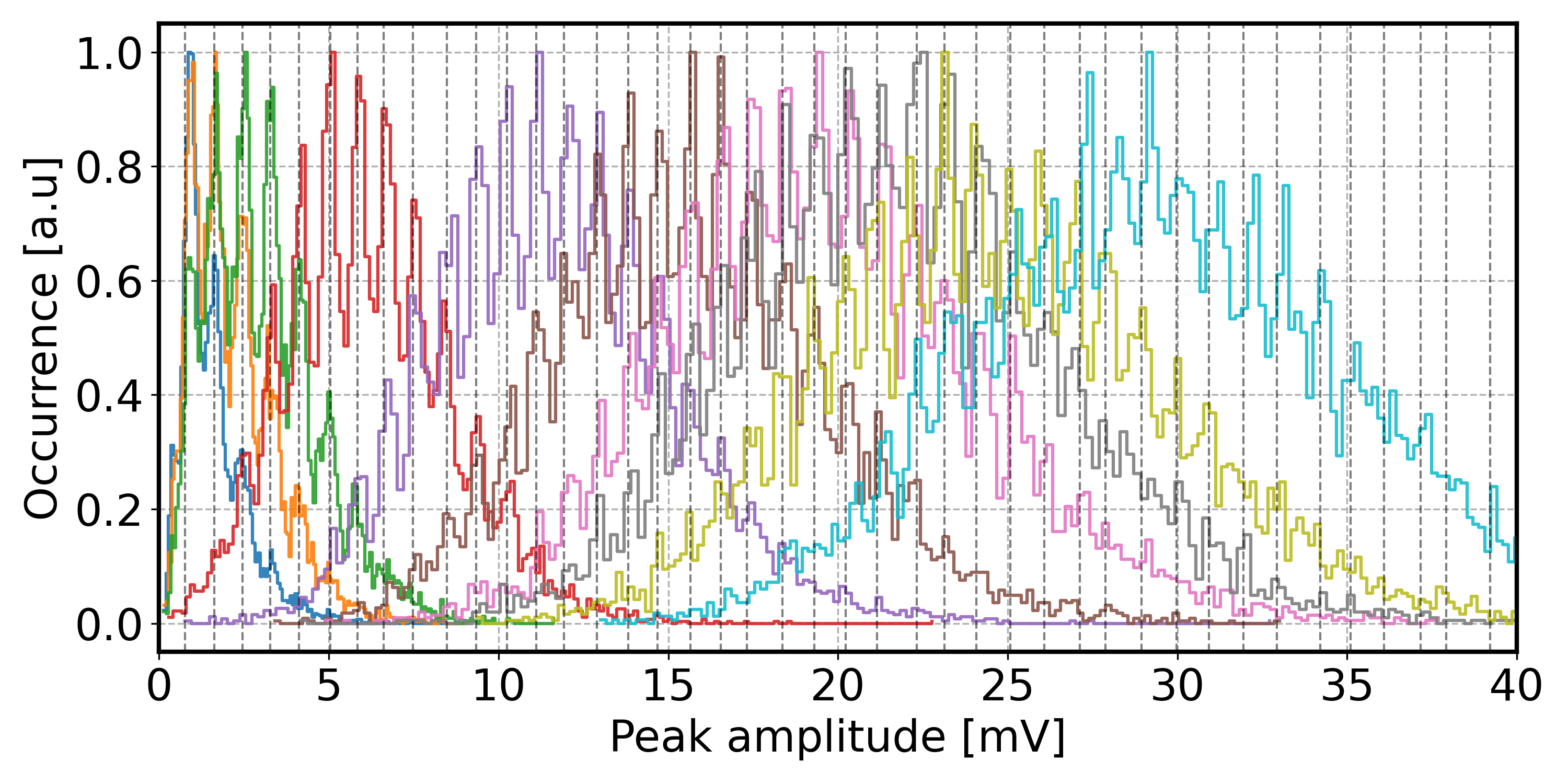}
    \caption{Overlaid amplitude histograms for different source intensities, each derived from 10,000 acquisitions.}
    \label{fig:histograms}
\end{figure}

By varying the source intensity, a total of 55 distinct photoelectron peaks were identified across overlaid histograms (Fig.~\ref{fig:histograms}). The system response remains linear with no evidence of saturation in the range of interest.

\subsection{Discussion}
The measured performance shows that the integrated sensor and ASIC can preserve the fast pulse structure required for Cherenkov imaging while resolving single-photoelectron signals. The sub 4~ns FWHM response is compatible with nanosecond-scale timing measurements, and the clear separation of photoelectron peaks indicates low-noise operation at low light levels. The on-chip summing of the four sensor quarters retains the benefits of segmentation while reducing the number of external readout channels. 
Compared with the existing PMT camera \cite{PACTA}, this work achieves a $4\times$ gain in spatial resolution, reaching 0.05 degrees per pixel, lowers the energy sensitivity to approximately 10 GeV and reduces the front-end power consumption by a factor $5$ when accounting for the same sensing area. Together, these results show that the proposed architecture improves detector granularity, threshold performance, and power consumption relative to the current camera design.
Further validation of channel-to-channel uniformity, long-term stability, and resolution under night-sky background rates up to 1~GHz per pixel will be needed for full camera deployment.

\subsection{Conclusions}
We have developed and characterized a co-designed SiPM sensor and front-end ASIC detector system achieving single-photoelectron resolution, 4~ns FWHM timing, and linear response from 1 to 130 photoelectrons. The presented system meets the stringent requirements for IACT cameras and positions itself as a viable candidate for the upgrade of the Large-Sized Telescope in the Cherenkov Telescope Array Observatory providing improved performance.


\begin{thebibliography}{00}

\bibitem{CTA} Cta Consortium. Science with the Cherenkov Telescope Array. World Scientific, 2018.
\bibitem{FACT} Biland, A., et al. "Calibration and performance of the photon sensor response of FACT—the first G-APD Cherenkov telescope." JINST 9.10 (2014): P10012.
\bibitem{CITIROC} Fleury, J., et al. "Petiroc and Citiroc: front-end ASICs for SiPM read-out and ToF applications." JINST 9.01 (2014): C01049-C01049.
\bibitem{TOFPET} Bugalho, R., et al. "Experimental characterization of the TOFPET2 ASIC." JINST 14.03 (2019): P03029-P03029.
\bibitem{TARGET} Albert, A., et al. "TARGET 5: A new multi-channel digitizer with triggering capabilities for gamma-ray atmospheric Cherenkov telescopes." Astroparticle Physics 92 (2017): 49-61.

\bibitem{FANSIC} Giangrande, L., et al. "FANSIC: A Fast ANalog SiPM Interface Circuit for the readout of large silicon photomultipliers." NIM-A, 1077 (2025): 170523.
\bibitem{PACTA} Sanuy, A., et al. "Wideband (500 MHz) 16 bit dynamic range current mode PreAmplifier for the CTA cameras (PACTA)." JINST 7.01 (2012): C01100-C01100.

\end{thebibliography}
\end{document}